\documentclass[aps,twocolumn,preprintnumbers,groupedaddress,superscriptaddress,floatfix,
tightenlines,reprint,nofootinbib,longbibliography]{revtex4-1}
\usepackage{mathrsfs}
\usepackage{natbib}
\usepackage{verbatim}
\usepackage{enumerate}
\usepackage{graphicx,bbm,amsbsy,amsfonts,amssymb,amsmath}
\usepackage{bm}
\usepackage{hyperref}
\usepackage{booktabs}
\usepackage{gensymb}

\usepackage{footmisc}

\usepackage{blindtext}

\def\ZZ{\mathbb{Z}} 
\def\RR{\mathbb{R}}

\def\2{{(2)}}
\def\1{{(1)}}
\def\0{{(0)}}
\def\m1{{(-1)}}

\let\oldAA\AA
\renewcommand{\AA}{\text{\normalfont\oldAA}}

\hypersetup{
    colorlinks=true,       
    linkcolor=red,          
    citecolor=blue,        
    filecolor=magenta,      
    urlcolor=blue           
}

\newcommand{\beq}{\begin{eqnarray}}
\newcommand{\eeq}{\end{eqnarray}}
\begin{document}

\title{Emergent 1-form symmetries}
\author{Aleksey Cherman}
\email{acherman@umn.edu}
\affiliation{School of Physics and Astronomy, University of Minnesota, Minneapolis, MN 55455}
\author{Theodore Jacobson}
\email{tjacobson@physics.ucla.edu}
\altaffiliation{Current affiliation: Mani L. Bhaumik Institute for Theoretical Physics, Department of Physics and Astronomy, University of California, Los Angeles, CA 90095, USA}
\affiliation{School of Physics and Astronomy, University of Minnesota, Minneapolis, MN 55455}

\begin{abstract}
We explore the necessary conditions for $1$-form symmetries to emerge in the
long-distance limit when they are explicitly broken at short distances. A minimal requirement is that there exist operators which become topological at long distances \emph{and} that these operators have non-trivial correlation functions. These criteria are obeyed when the would-be emergent symmetry is spontaneously broken, or is involved in 't Hooft anomalies. On the other hand, confinement, i.e. a phase with unbroken 1-form symmetry, is nearly incompatible with the emergence of 1-form symmetries.   We comment on some implications of our results for QCD as well as the idea of Higgs-confinement continuity.
\end{abstract}

\maketitle
\flushbottom

\section{ Introduction}  

The global symmetries we encounter in nature are generally explicitly broken at short distances, 
but they often emerge in long-distance QFT descriptions.  In recent
years, it has become appreciated that relativistic QFTs can have generalized
global symmetries~\cite{Gaiotto:2014kfa}, such as higher-form symmetries that
act on non-local operators like lines and surfaces.  For example, $1$-form
symmetries can act on Wilson loops in gauge theories, and have led to a helpful
new perspective on color confinement, among many other
results~\cite{McGreevy:2022oyu,Cordova:2022ruw}. In this paper we explore how
$1$-form symmetries can emerge in long-distance limits even when they are
explicitly broken at short distances.

Standard `0-form' symmetries act on local operators and can be explicitly broken
by adding an appropriate charged local operator to a Lagrangian density. The
fate of the symmetry at long distances is then determined by the scaling
dimension of the perturbing operator.  It is difficult (though not impossible,
see Ref.~\cite{Iqbal:2021rkn}) to use similar renormalization-group-style ideas to
study the long-distance fate of explicitly broken $1$-form symmetries, because
the charged objects are line operators.  In practice, explicit breaking of
one-form symmetries in gauge theories arises due to couplings to dynamical
charged matter fields~\cite{Gaiotto:2014kfa}.  Then there are two basic ways a
$1$-form symmetry can emerge. First, it can appear upon taking a limit in the
space of QFTs. If we consider a sequence of field theories where the mass $m$ of
the symmetry-breaking fields increases with all other physical parameters held
fixed, then a $1$-form symmetry must certainly emerge in the limit $m\to
\infty$. Second, a $1$-form symmetry might emerge in a \emph{fixed} QFT (in this
context, with fixed $m$) in the long-distance limit. How and when this happens
is more subtle, and such `infrared-emergent $1$-form symmetries' will be our
main focus in this paper.

Below we will state a condition that must be satisfied for a QFT to have a
non-trivial emergent $1$-form symmetry in the long-distance limit.  We then
analyze how this condition is satisfied (or not) in some simple examples. Along
the way we will make contact with other discussions of emergent and approximate
$1$-form symmetries in the recent
literature~\cite{Delacretaz:2019brr,Iqbal:2021rkn,Cordova:2022rer,Pace:2022cnh,Pace:2023gye,Armas:2023tyx,Cherman:2022eml} and discuss the role of 't Hooft anomalies in
the emergence of $1$-form symmetries. We also explain why there is no emergent $1$-form symmetry in QCD even when all of the
quarks have large finite masses, and implications for Higgs-confinement continuity with fundamental matter.

\section{ Exact $1$-form symmetry} A modern definition of exact symmetries in
relativistic QFTs is via the existence of topological operators along with data
on how they act in correlation functions~\cite{Gaiotto:2014kfa}. These operators
can be thought of as charges which generate the symmetry. We focus on invertible
$1$-form symmetries in relativistic QFTs in $d$-dimensional Euclidean spacetime.  The associated
topological operators $U_\alpha(M_{d-2})$ live on codimension-$2$ manifolds
$M_{d-2}$ and are labelled by an element $\alpha$ of the symmetry group $G$.
Their correlation functions have a purely topological dependence on $M_{d-2}$,
and they satisfy fusion rules given by group composition in $G$. The charged
operators are line operators $W_R(C)$ where $C$ is a closed loop and $R$
contains information about the charge of the operator.\footnote{Some $3+1$d systems have more exotic topological surface operators that only act on other surface operators, not line operators, see e.g.~\cite{Hsin:2019fhf}.  Our goal here is to understand more conventional
$1$-form symmetries that act on line operators.} For example, in $U(1)$ Maxwell
theory with a $1$-form gauge field $a = a_{\mu} dx^{\mu}$, there is a $U(1)$
$1$-form `electric' symmetry generated by Gukov-Witten operators which prescribe
the holonomy of $a$ on infinitesimal circles $C$ linking $M_{d-2}$ to be
$e^{i\alpha}$. In this simple example one can also give the less abstract
definition
\begin{align}
    U_{\alpha}(M_{d-2}) \sim e^{i \alpha \int_{M_{d-2}} \star j } 
\end{align}
where $j =  -\frac{i}{g^2} da$ is the electric $2$-form conserved current, $d$
is the exterior derivative, $\star$ is the Hodge star operator, and $g$ is the
gauge coupling.  

The
charged objects are electric Wilson loops $W_n(C) = \exp{\left(i n \oint_C
a\right)}$, and 
\begin{align}
\frac{\langle U_{\alpha}(M_{d-2}) W_n(C)\rangle}{\langle U_{\alpha}(M_{d-2})\rangle \langle W_n(C)\rangle } = 
e^{in\alpha\, \textrm{Link}(M_{d-2},C)} 
\label{eq:exact_UW}
\end{align}

As with any exact group-like symmetry, invertible $1$-form symmetries lead to
selection rules on finite-volume correlation functions~\cite{Gaiotto:2014kfa}.  In the infinite-volume limit they can be
spontaneously broken, which is signaled by a perimeter-law behavior for Wilson
loops, and is interpreted as deconfinement of test charges. For example  in 4d QED $ \langle W_1(C)\rangle \to e^{-\mu P(C)}$ when
$C$ is large, $P(C)$ is the perimeter of $C$ and $\mu$ is an energy scale that
depends on the choice of renormalization scheme. One can choose a counter-term
localized on $C$ to set $\mu = 0$, and then $ \langle W_1(C)\rangle \not=0$ for large Wilson
loops. If instead $ \langle W_1(C)\rangle$ goes to zero faster than a perimeter law for
large $C$, then $\lim_{|C| \to\infty} \langle W_1(C)\rangle = 0$ regardless of the choice of
scheme, and the symmetry is not spontaneously broken.  This signals charge confinement.

\section{ Emergent $1$-form symmetry} 

If a QFT contains dynamical minimal-charge
matter fields, then there are no topological codimension-$2$ operators that
satisfy~Eq.~\eqref{eq:exact_UW}, and hence no exact $1$-form
symmetry. The absence of such operators
can be established from the existence of non-trivial open Wilson lines. The
issue is that different ways of `shrinking' a symmetry generator in
the presence of an open Wilson line give manifestly different results, which at
the same time must be identical if the generator is topological~\cite{Rudelius:2020orz}.

Heuristically, an infrared-emergent $1$-form symmetry should be associated with
the existence of operators $U_{\alpha}(\Sigma_{d-2})$ which are only topological
in the long distance limit.   Then the existence of open Wilson lines is not an
issue, since the $U_{\alpha}(\Sigma_{d-2})$ operators only behave topologically
when they are large, and cannot be freely shrunk on open lines. 

We define an infrared-emergent $1$-form symmetry as
\begin{enumerate}[(ES)]
\item Existence of a set of operators $U_{\alpha}(\Sigma_{d-2})$ defined on codimension-$2$
manifolds $\Sigma_{d-2}$ with correlation functions that are topological and non-trivial in the
long-distance limit.    
\end{enumerate}
This differs from the definition of an exact $1$-form symmetry in three ways: it
involves a long-distance limit, does not explicitly refer to the action of
$U_{\alpha}(\Sigma_{d-2})$ on line operators, and does not assume that
$\Sigma_{d-2}$ is closed.  The nature of the correlation functions of $U_{\alpha}$ that remain
non-trivial in the long-distance limit can be quite subtle, and they do not always
involve genuine line operators~\cite{Kapustin:2014gua} or $U_{\alpha}$ operators defined on closed manifolds.


In what follows we explore the emergence of 1-form symmetries in a number of
simple examples, which are all variants of 3d scalar QED. Specifically, we consider parity-invariant $U(1)$ gauge
theory with or without magnetic monopoles coupled to matter fields with various
charges. We will see that whether our
definition of emergent $1$-form symmetry (ES) is satisfied can depend on the
realization of the symmetry when we consider a limit in the space of theories
where it becomes exact, and whether the symmetry is involved in 't Hooft anomalies in that limit. 

\section{Confinement versus emergence }  

Consider scalar QED with a charge-$1$ scalar field $\phi$ with mass $m$, 
\begin{align}
S = \int_{M_3} \left(\frac{1}{2g^2} (da)^2 + |d\phi- i a \phi|^2 + \ldots \right) \,,
\label{eq:linear_conf_action}
\end{align}
with a short-distance definition that allows
finite-action magnetic monopole-instantons with unit charge to contribute to the
path integral. The model has Gukov-Witten operators $U_{\alpha}(\Gamma)$ whose correlators depend on the geometry of $\Gamma$ at
finite $m$, but  become topological in the limit $m\to \infty$, where they
generate an exact $U(1)$ $1$-form global symmetry.  Monopole-instantons generate
a mass gap, and when $m\to \infty$ the theory is confining: large Wilson loops
have area-law fall-off with a string tension $\sigma$~\cite{Polyakov:1976fu}, so
the $1$-form symmetry is not spontaneously broken in this limit.  

If $m^2$ is finite and $m^2 \gg g^4$, it is natural to integrate out
$\phi$ to describe physics at distances large compared to $1/m$, leading to
the long-distance effective action
\begin{align}
S_{\rm eff} = \int_{M_3}\left( \frac{1}{2 g^2} (da)^2 +  \frac{c_4}{m^5} da^4 + \cdots \right)
\label{eq:Seff3dQED}
\end{align}
where $c_4 \sim \mathcal{O}(1)$ and $\cdots$ represents higher-order terms
containing higher powers of $d a$ and its derivative.  Note that a Chern-Simons
term cannot appear given parity invariance, and thus all terms which can appear
in $S_{\rm eff}$ are invariant under shifts of $a$ by a closed $1$-form
$\lambda$. One might therefore think that there is a robust infared-emergent $1$-form
symmetry when $m^2$ is finite and $m^2 \gg g^4$.  This turns out not to be true.

The key point is that the effective action in Eq.~\eqref{eq:Seff3dQED} is not actually
`effective' for all long-distance observables. While it is effective for
calculating correlation functions of widely-separated local operators, it fails
to correctly describe the correlation functions of large Wilson loops.  In particular,
consider $\langle W(C)\rangle$, $\langle U_{\alpha}(\Gamma) \rangle$, and $\langle U_{\alpha}(\Gamma)
W(C)\rangle$. The path integral over $\phi$ produces a formal sum over all
possible insertions of `dynamical' minimal-charge Wilson loops, so that
schematically 
\begin{subequations}
  \begin{align}
    &\langle U_{\alpha}(\Gamma) \rangle \sim 
    \sum_{C'} e^{-\mu P(C')} \langle U_{\alpha}(\Gamma) W(C') \rangle_{0} + \cdots
     \label{eq:U_screening} \\ 
    &\langle W(C) \rangle \sim \sum_{C'}e^{-\mu P(C')} \langle  W(C) W(C') \rangle_{0} + \cdots
     \label{eq:W_screening} \\
   &\langle U_{\alpha}(\Gamma) W(C) \rangle     \sim \nonumber\\
    &\sum_{C'} e^{-\mu P(C')} \langle U_{\alpha}(\Gamma)W(C) W(C')\rangle_{0} + \cdots  
    \label{eq:UW_screening}
  \end{align}
  \label{eq:screening_correlators}
\end{subequations}
where $\langle \cdot\rangle_{0}$ denotes a (connected) correlator evaluated in the pure gauge
theory, $\mu$ is some appropriate mass scale (such as the mass of $\phi$), and
$\cdots$ represents terms with multiple dynamical Wilson loops. The above
representation of expectation values arises naturally  in the large mass (hopping)
expansion on the lattice~\cite{Montvay:1994cy}, or in the worldline formalism in
the
continuum~\cite{Feynman:1950ir,Feynman:1951gn,Bern:1990cu,Bern:1991aq,Strassler:1992zr}.

\begin{figure}[h]
  \includegraphics[width=0.8\columnwidth]{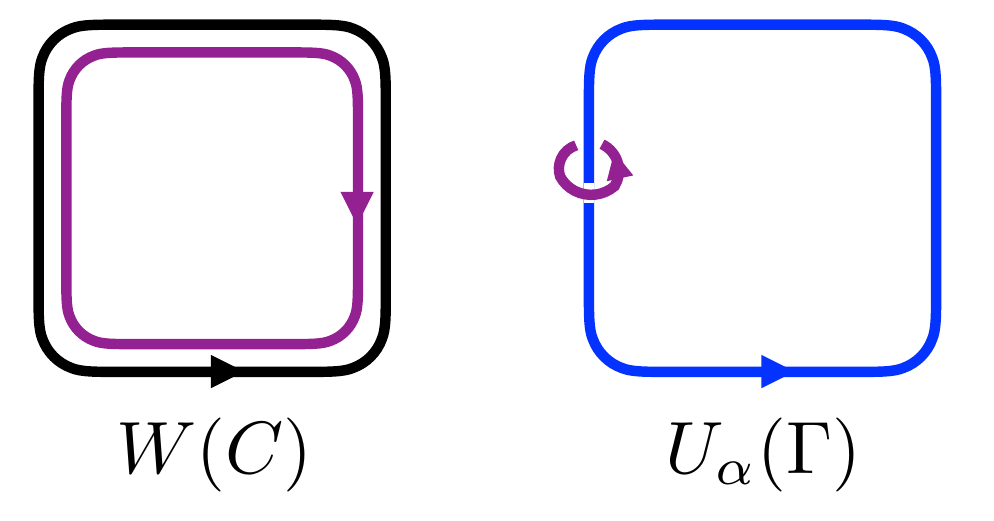}
  \caption{Left: In a confining theory coupled to heavy matter, the expectation value of a large Wilson loop $W(C)$ (black
   curve) is dominated by contributions that include a large screening `matter'
   Wilson loop (purple curve) arising from integrating out a charged matter
   field. Right: The expectation value of $U_{\alpha}(\Gamma)$ (blue curve) is
   dominated by the contributions of small Wilson loops (purple curve) arising
   from integrating out the charged matter field.}
  \label{fig:matter_loops}
\end{figure}

We start by considering the one-point function of the Gukov-Witten operator in
Eq.~\eqref{eq:U_screening}.  The fact that large dynamical Wilson loops $W(C')$
come with a suppression factor $e^{-\mu P(C')}$ implies that for large $\Gamma$
the sum in Eq.~\eqref{eq:U_screening} is dominated by small loops $C' \sim 0$.
Summing over Wilson loop insertions on small curves $C'$ that link $\Gamma$
(illustrated on the right-hand side of Fig.~\ref{fig:matter_loops}) generates
perimeter-law behavior for the Gukov-Witten operators 
\begin{align}
\langle U_{\alpha}(\Gamma)\rangle \sim e^{-\mu_{\alpha}P(\Gamma)} 
\,,
\label{eq:U_perimeter}
\end{align}
where $\mu_{\alpha}$
is a non-universal mass scale that depends on $\alpha$ as well as e.g. the mass
of the charged particle. But we can set $\mu_{\alpha}$ to zero by a counter-term
localized on $\Gamma$, producing operators which are topological for large
$\Gamma$ so long as $m^2 \gtrsim g^4$.   This is consistent with a naive
analysis based on the effective field theory Eq.~\eqref{eq:Seff3dQED}.  

We now consider Eqs.~\eqref{eq:W_screening}, \eqref{eq:UW_screening}.  
Since $\langle W(C) \rangle_{0}$ has confining behavior, there is a competition between the perimeter-suppression $e^{-\mu P(C')}$ of dynamical Wilson loops $W(C')$ and the more severe (in this case area-law) suppression of the probe Wilson loop $W(C)$ in the pure gauge theory. As a result, for large contours $C$, large dynamical Wilson loops are favored rather than suppressed, and  $\langle W(C) \rangle$ is dominated by fluctuations around the `screening
loop' $C' = \overline{C}$ running opposite to $C$. This is illustrated in the
left-hand side of Fig.~\ref{fig:matter_loops}. Such contributions go to zero
with the perimeter of $C$ rather than its area.  This is simply the
standard physics of screening, which is not captured in the local effective
field theory of Eq.~\eqref{eq:Seff3dQED}! The same conclusion holds for the
correlator of the Wilson loop and Gukov-Witten operator in
Eq.~\eqref{eq:UW_screening}. Again, the dominant contribution involves a
screening loop, and as a result the Gukov-Witten operator measures zero charge.

To summarize, the fact that large Wilson loops are screened
implies that their charge cannot be detected at long distances. Instead, 
\begin{align}
\frac{\langle U_{\alpha}(\Gamma) W(C)\rangle}{{\langle U_{\alpha}(\Gamma)\rangle \langle  W(C)\rangle}} 
= 1 
\label{eq:trivial_correlator}
\end{align} 
for well-separated curves $\Gamma, C$ which tend to infinity.  Effectively,
confinement in the limit $m\to \infty$ (meaning that $\langle W(C) \rangle \to
0$ faster than a perimeter law in that limit) implies that when $m$ is finite
all Wilson loops flow to the identity line operator at long distances,
leaving nothing for the would-be emergent $1$-form symmetry to act on.  As a
result the correlation functions of $U_{\alpha}(\Gamma)$ are trivial --- all
charges are screened. Therefore our emergent-symmetry definition (ES) is not
satisfied, and there is no emergent $1$-form symmetry at finite positive $m^2$.
Finally,  if $-m^2 \gtrsim g^4$, screening loops due to $\phi$ proliferate.  Then
there is no approximation in which we can neglect the matter field, and so there is also no emergent $1$-form symmetry for  $-m^2 \gtrsim g^4$. So there is no reason to
expect an emergent $1$-form symmetry for \emph{any} finite $m^2$ in this model.  

We conclude that confinement is almost entirely incompatible with the emergence
of $1$-form symmetries.  For example, there is no infrared-emergent $1$-form
symmetry in 4d QCD with fundamental quarks of any finite mass $m_q$, no matter
how large $m_q$ is compared to the strong scale.

\section{Deconfinement and emergence}

We now discuss what happens when an exact
$1$-form symmetry is spontaneously broken, and then also explicitly broken by
the addition of heavy charged matter fields. To this end, we couple 3d $U(1)$
gauge theory to a charge $N$ scalar $\chi$,  
\begin{equation}
S = \int_{M_3}
\left(\frac{1}{2g^2} (da)^2 + |d\chi- i N a \chi|^2 + \ldots \right)\,,
\end{equation} 
and assume $\chi$ is the only electrically-charged field. This theory
has an exact $\ZZ_N$ 1-form symmetry generated by Gukov-Witten operators
$U_{k}(\Gamma)$, defined to have holonomies $e^{\frac{2\pi i k}{N}}$ on
infinitesimal contours linking $\Gamma$. We condense $\chi$, Higgsing the gauge
group to $\ZZ_N$. In this phase large Wilson loops have a perimeter-law scaling, and the
$1$-form symmetry is spontaneously broken.

%

We now put in a unit-charge scalar $\phi$ with mass $m$ to explicitly break the
$1$-form symmetry.  When  $m^2 \gtrsim g^4$ the $1$-form symmetry emerges in the
long-distance limit with $m$ held fixed. To see this we examine the
behavior of the correlation functions in Eq.~\eqref{eq:screening_correlators} resulting
from formally summing over $\phi$ configurations, with $\langle \cdot \rangle_{0}$
now interpreted as correlation functions of the theory without $\phi$. The
analysis of $\langle U_{\alpha}(\Gamma) \rangle$ goes through as before, but the
perimeter-law scaling of $\langle W(C) \rangle_{0}$ now implies that small matter
loops (as opposed to screening loops with $C' \sim \overline{C}$) dominate
$\langle U_{\alpha}(\Gamma) W(C) \cdots \rangle$ when $C$ is very large, and
\begin{align}
\frac{\langle U_{k}(\Gamma) W(C)\rangle}{  \langle  U_{k} (\Gamma)\rangle \langle  W(C)\rangle} =  e^{\frac{2\pi i k}{N} \textrm{Link}(\Gamma,C)} 
\end{align}
for very large and well-separated loops $C, \Gamma$.  Therefore we have a set of
operators $U_{\alpha}(\Gamma)$ which have non-trivial topological correlation
functions in the long-distance limit, our definition (ES) is satisfied, and there is an infrared emergent
$\mathbb{Z}_N$ $1$-form symmetry so long as $\chi$ is condensed and $\phi$ is
not.  

This analysis resonates with recent discussions of emergent $1$-form symmetries
in
Refs.~\cite{Iqbal:2021rkn,Cordova:2022rer,Pace:2022cnh,Pace:2023gye,Armas:2023tyx}
which assumed (explicitly or implicitly) that they were working in the
spontaneously-broken situation discussed in this section. In fact, there is a
standard way to understand the above result without invoking the modern language
of higher-form symmetries. The charge-$N$ abelian Higgs model flows to a $\ZZ_N$
gauge theory at long distances, described in the continuum by a BF action
$S_{\rm BF} = \frac{i N}{2\pi} \int_{M_3} b \wedge da$ where $b$ is a (emergent)
$U(1)$ $1$-form gauge field, see e.g. Ref.~\cite{Hansson:2004wca}. This is a
topological field theory with $2^N$ ground states on a large spatial torus, and
it has long been known (see~\cite{Wen:1991rp,Vestergren_2005,Hastings_2005})
that these ground states remain degenerate (and the long-distance  BF
description remains valid) after the addition of unit-charge matter so long as
it does not `condense.'  From the more modern perspective, BF theory can be
thought of as an effective field theory for a spontaneously-broken
$\mathbb{Z}_N$ $1$-form symmetry~\cite{Gaiotto:2014kfa}.

\section{'t Hooft anomaly and charged operators}  
\label{sec:anomaly}

We now explore the interplay between 't Hooft anomalies and emergent $1$-form
symmetries. We again start with $U(1)$ gauge theory in 3d, $S = \int_{M_3}
\frac{1}{2g^2} (da)^2$, but this time we do \emph{not} allow dynamical magnetic
monopole-instantons in the UV completion. This gives rise to a 0-form `magnetic'
symmetry $U(1)^{(0)}_m$ associated with the conserved current $j_m^{(1)}
=\frac{1}{2\pi} \star da$. This symmetry is generated by topological surface
operators $V_{\beta}(\Sigma)$ which act on point-like charge $k$ monopole
operators $e^{i k\sigma}$ where $\sigma$ is the $2\pi$-periodic scalar field
dual to $a$. As before, the $U(1)^{(1)}_e$ electric symmetry is generated by
topological line operators $U_{\alpha}(\Gamma)$ which act on charge $q$ Wilson
lines $W_{q}(C)$.  

The $U(1)^{(0)}_m$ and $U(1)^{(1)}_e$ symmetries have a mixed 't Hooft anomaly, see e.g. \cite{Gaiotto:2014kfa}.
This can be detected by turning on $1$-form and $2$-form $U(1)$ background
gauge fields $A_m$ and $B_e$, with $0$-form and $1$-form background gauge transformations
$A_m \to A_m + d\Lambda_m, B_e \to B_e + d\Lambda_e$, $a\to a + \Lambda_e$, so that 
\begin{align}
\label{eq:naive_anomaly}
S[A,B] = \int_{M_3} \bigg[ &\frac{1}{2g^2} (da - B_e)^2 + \frac{i}{2\pi} A_m \wedge da \\
& - \frac{i}{2\pi}
 b_{\rm CT} A_m \wedge B_e + \cdots \bigg] \nonumber
\end{align}
where $\cdots$
stands for other background-field counter-terms in addition to the $A_m\wedge B_e$ term.  
While it may seem from Eq.~\eqref{eq:naive_anomaly} one can preserve background
gauge invariance for $U(1)^{(0)}_m$ by setting $b_{\rm CT} = 0$ (or do the same
for $U(1)^{(1)}_e$ by setting $b_{\rm CT} = 1$), the situation is more subtle
and is detailed in Appendix \ref{sec:anomaly_done_right}. In any case there is no choice of
counterterms which can enforce background gauge invariance for \emph{both} $A_m$
and $B_e$ simultaneously, indicating an 't Hooft anomaly.

\begin{figure}[h]
  \includegraphics[width=0.9\columnwidth]{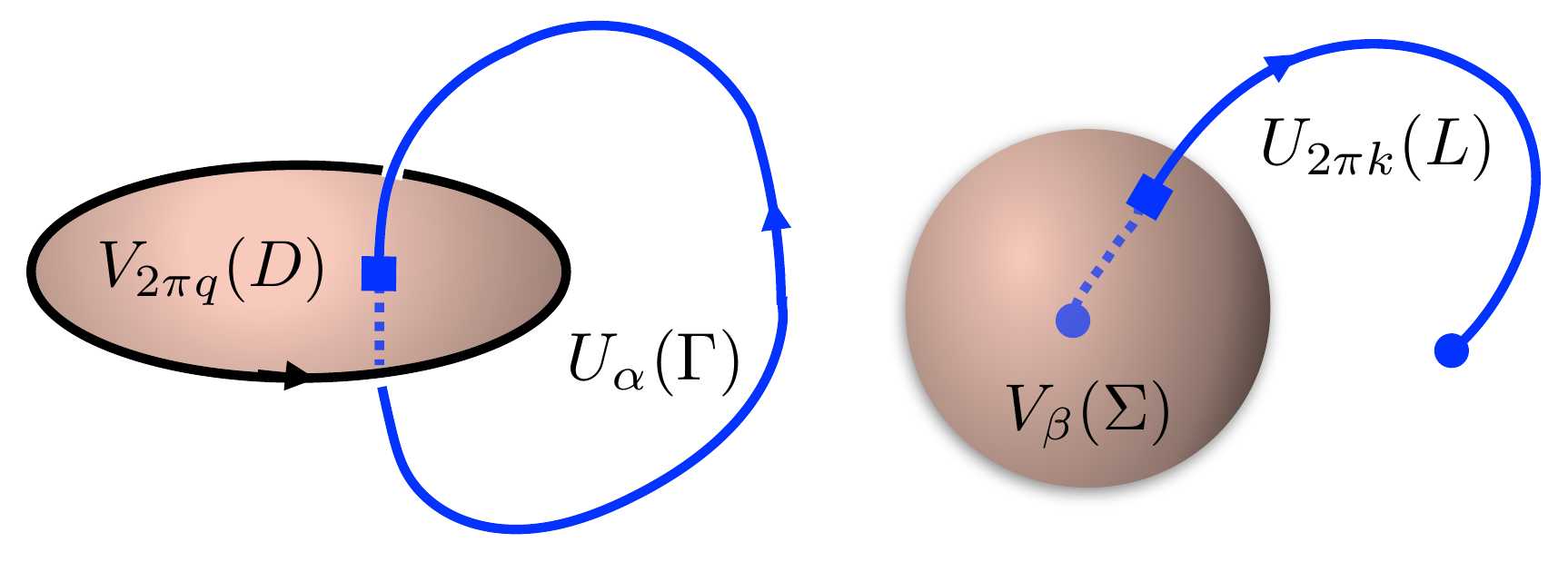}
  \caption{The 't Hooft anomaly of 3d pure $U(1)$ gauge theory with a $U(1)^{(0)}_m
   \times U(1)^{(1)}_e$ symmetry implies that certain unusual correlation
   functions of the symmetry generators are non-trivial. These correlation
   functions involve operator intersections (denoted by squares in the figure)
   and are sensitive to the choice of counter-terms.  But it is not possible to
   fully trivialize the expectation value of the combination of operators given
   in Eq.~\eqref{eq:non_trivial_correlator}, which is illustrated in this figure.}
  \label{fig:anomaly_correlators}
\end{figure}

The 't Hooft anomaly implies the following correlation functions:
\begin{subequations}
\label{eq:anomaly_correlators}
\begin{align}
    \frac{\langle V_{2\pi q }(D)\, U_{\beta}(\Gamma) \rangle}
    {\langle V_{2\pi q}(D)\rangle \langle U_{\beta}(\Gamma) \rangle} 
    &= e^{i q\beta \,(1-b_{\rm CT})\, \textrm{Link}(\partial D, \Gamma)} \,,
    \label{eq:surface_anomaly} \\
     \frac{\langle V_{\alpha}(\Sigma)\, U_{2\pi k}(L) \rangle}
    {\langle V_{\alpha}(\Sigma)\rangle \langle U_{2\pi k}(L) \rangle} 
    &= e^{ik\alpha \, b_{\rm CT} \textrm{Link}(\Sigma, \partial L)} \,,
    \label{eq:endpoint_anomaly}
\end{align}
\end{subequations}
where $q, k \in \mathbb{Z}$, and $D$ and $\Sigma$ are (respectively) open and
closed surfaces, while $L$ and $\Gamma$ are open and closed lines, which are
sketched in Fig.~\ref{fig:anomaly_correlators}.  We note that an
open surface operator can be modified by attaching an arbitrary
line operator to its boundary, with similar remarks holding for open line operators. Our choice in
Eq.~\eqref{eq:anomaly_correlators} is such that together, the bulk and boundary
of the operators $V_{2\pi q}(D), U_{2\pi k}(L)$ are topological. 

Absent the anomaly, $V_{2\pi q}(D)$ would be a trivial surface operator which
contributes only contact terms to correlation functions which can all
be trivialized by a choice of local counter-terms. But if we take $b_{\rm CT} =
0$, Eq.~\eqref{eq:surface_anomaly} tells us that the boundary of $V_{2\pi q}(D)$
acts like a charge $q$ \emph{topological} Wilson line. By comparison, the
genuine Wilson lines $W_{q}(C)$, which are not attached to surfaces, are not
topological in this model --- they are (logarithmically) confined! Similar remarks hold for
Eq.~\eqref{eq:endpoint_anomaly}, which says that the end-points of $U_{2\pi
k}(L)$ behave like a pair of charge $\pm k$ monopole operators (when $b_{\rm CT}
= 1)$, with only a topological dependence on $L$. In summary,
Eq.~\eqref{eq:anomaly_correlators} tells us that in a given scheme the 't Hooft
anomaly adds $ V_{2\pi q }(D)$ or $U_{2\pi k }(L)$ to the list of distinct
operators charged under $U(1)^{(0)}_m$ and $U(1)^{(1)}_e$, which would otherwise
have only included $W_{q}(C)$ and $e^{i k \sigma}$. The scheme-independent
statement is that at least one of the correlators in
Eq.~\eqref{eq:anomaly_correlators} is nontrivial. Equivalently, the following
correlator is nontrivial in any scheme:
\begin{multline}
\frac{\langle V_{2\pi q}(D) U_\beta (\Gamma)\, V_{\alpha}(\Sigma) U_{2\pi k}(L) \rangle}
{\langle V_{2\pi q}(D)\rangle \langle U_\beta (\Gamma)\rangle
\langle V_{\alpha}(\Sigma)\rangle\langle U_{2\pi k}(L) \rangle } \\
= e^{ik\alpha \, b_{\rm CT} \textrm{Link}(\Sigma, \partial L)} 
\,e^{i q\beta \,(1-b_{\rm CT})\, \textrm{Link}(\partial D, \Gamma)} \not =1.
\label{eq:non_trivial_correlator}
\end{multline}

The quick way to see why Eq.~\eqref{eq:anomaly_correlators} is true is to consider
the effects of singular background gauge transformations. An insertion of
$V_{2\pi q}(D)$ is equivalent to turning on the background $A_m =2\pi q
\delta^{(1)}(D)$, which can be removed by a singular gauge transformation
$\Lambda_A$ which winds by $-2\pi q$ around $\partial D$. Similarly, an
insertion of $U_{\beta}(\Gamma)$ with $\Gamma$ contractible is equivalent to
taking $B_e = \beta \delta^{(2)}(\Gamma)$, which can be removed by the
background gauge transformation by $\Lambda_B = - \beta \delta^{(1)}(S)$, where
$\partial S = \Gamma$.  Doing these transformations in $S[A,B]$, we obtain
\begin{align}
    \delta S[A,B] &= \frac{i 2\pi q \beta (1-b_{\rm CT} )}{2\pi} \int_{M_3}  \delta^{(1)}(D) \wedge  \delta^{(2)}(\Gamma)
    \nonumber \\
    &=i q\beta (1-b_{\rm CT} )\, \textrm{Link}(\partial D, \Gamma) \, ,
\end{align}
reproducing Eq.~\eqref{eq:surface_anomaly}.  A very similar computation gives
Eq.~\eqref{eq:endpoint_anomaly}.  We also derive this result in Appendix
\ref{sec:anomaly_done_right} using the language of coordinate patches, cochains, and transition
functions for readers who are nervous about manipulations of singular gauge
transformations.  This analysis shows the precise sense in which the 't Hooft
anomaly remains non-trivial even in locally-flat background fields, which is not
obvious from its naive form in Eq.~\eqref{eq:naive_anomaly}.

There is a well-known argument that an 't Hooft anomaly implies that the ground
state cannot be trivially gapped.  If it were trivially gapped, one could
simultaneously gauge both $U(1)^{(0)}_m$ and $U(1)^{(1)}_e$ in the (empty)
long-distance EFT, which would be inconsistent with the 't Hooft anomaly.  The
same constraint on the ground state structure follows from
Eq.~\eqref{eq:non_trivial_correlator}.   If the ground state on $M_3 = \mathbb{R}^3$
were trivially gapped, then all correlation functions must either go to zero or unity in the long-distance limit.\footnote{This may require adjusting the coefficients of local counterterms. The more precise statement is that the long-distance limit of any correlation function in the trivially gapped theory reduces to a pure contact term which can be removed by an appropriate choice of scheme.}  But Eq.~\eqref{eq:non_trivial_correlator} is
non-trivial for any contractible $\Gamma$ and $\Sigma$, no matter how large, so
the ground state on $\mathbb{R}^3$ cannot be trivially gapped.    

\section{'t Hooft anomaly and emergence} 

Now we reintroduce our charge-$1$ scalar
$\phi$ with mass $m$, 
\begin{equation} \label{eq:charge1higgs} 
S = \int_{M_3} \left(\frac{1}{2g^2} (da)^2 + |d\phi- i a
\phi|^2 + m^2 |\phi|^2 + \ldots \right)\,,
\end{equation} 
so that $U(1)^{(1)}_e$ is explicitly
broken for any finite $m^2$.  However, in contrast to the analysis after
Eq.~\eqref{eq:linear_conf_action}, we continue to assume that there are no dynamical
monopole-instantons, so that $U(1)^{(0)}_m$ is preserved.   The explicit
breaking of $U(1)^{(1)}_e$ means that the 't Hooft anomaly seems to be gone, as
it no longer makes sense to consider (background) gauging both symmetries.
However, is there a range of microscopic parameters for which this QFT has an
emergent $1$-form symmetry in the long-distance limit?

This question was considered from the particle-vortex dual perspective
\cite{Peskin:1977kp,Dasgupta:1981zz,Karch:2016sxi}  in
Ref.~\cite{Delacretaz:2019brr}. The authors of Ref.~\cite{Delacretaz:2019brr}
argued that the 't Hooft anomaly of the Nambu-Goldstone effective field theory for a $2+1$d
superfluid, which is dual to our $S_{\rm eff}$ from Eq.~\eqref{eq:Seff3dQED},
implies the existence of a robust gapless mode. Here we are posing the question
with regard to our definition (ES), which requires us to analyze the correlation
functions of would-be topological operators. 

When $m^2$ is sufficiently positive, the discussion
following~Eq.~\eqref{eq:screening_correlators} implies that we can construct
approximately-topological operators $U_{\beta}(\Gamma)$ and $U_{2\pi k}(L)$,
with $\Gamma$ (resp. $L$) a closed (resp. open) line. However, $\langle W_q(C) \rangle \to
0$ faster than a perimeter law when $m^2 \to \infty$, so $W_q(C)$ flows to the
trivial operator for large $C$ for any finite $m$. So one may worry that the
would-be $1$-form symmetry generators $U_{\beta}(\Gamma)$ have nothing to act on
in the long-distance limit.  

However, the situation is more interesting.  Since $U(1)^{(0)}_m$ is not
explicitly broken,  $V_{2\pi q }(D)$ and $V_{\beta}(\Sigma)$ are topological
surface operators.   We can now consider Eq.~\eqref{eq:naive_anomaly} with $B_e$
replaced by a $2$-form background field $B$, which can no longer be interpreted
as a background gauge field,  since $U(1)^{(1)}_e$ is explicitly broken.
Nevertheless, setting $B = \beta \delta^{(2)}(\Gamma)$ has the effect of
inserting a non-topological Gukov-Witten operator $U_{\beta}(\Gamma)$. For
large $\Gamma$, however, $U_{\beta}(\Gamma)$ is approximately topological. If
$b_{\rm CT} \neq 1$, $V_{2\pi q }(D)$ has non-trivial approximately-topological
correlation functions with $U_{\beta}(\Gamma)$ when $\Gamma$ is large.  Crucially, the fact that $V_{2\pi q }(D)$ is exactly topological for any $m^2$
means that it is not screened by charged matter loops. If $b_{\rm CT} = 1$, the
approximately topological operator $U_{2\pi k}(L)$ has non-trivial correlation
functions with the exactly topological operator $V_{\beta}(\Sigma)$. Therefore
Eq.~\eqref{eq:anomaly_correlators} remains non-trivial in the long-distance limit
for sufficiently positive fixed $m^2$. 

If $m^2$ is not sufficiently positive,
then matter Wilson loops proliferate, and $U_{\beta}(\Gamma), U_{\beta}(L)$ flow to either the zero or identity operators at long distances depending on the choice of counterterms.  

We thus see that our definition (ES) is satisfied when $m^2$ is sufficiently
positive, so that the theory described by Eq.~\eqref{eq:charge1higgs} has a
$U(1)^{(1)}_e$ emergent $1$-form symmetry, despite the fact that the
$U(1)^{(1)}_e$ symmetry is \emph{not} spontaneously broken in the limit $m\to
\infty$, where test charges are (logarithmically) confined.

\section{Consequences} 
\label{sec:consequences}

\begin{figure}[h]
  \includegraphics[width=0.9\columnwidth]{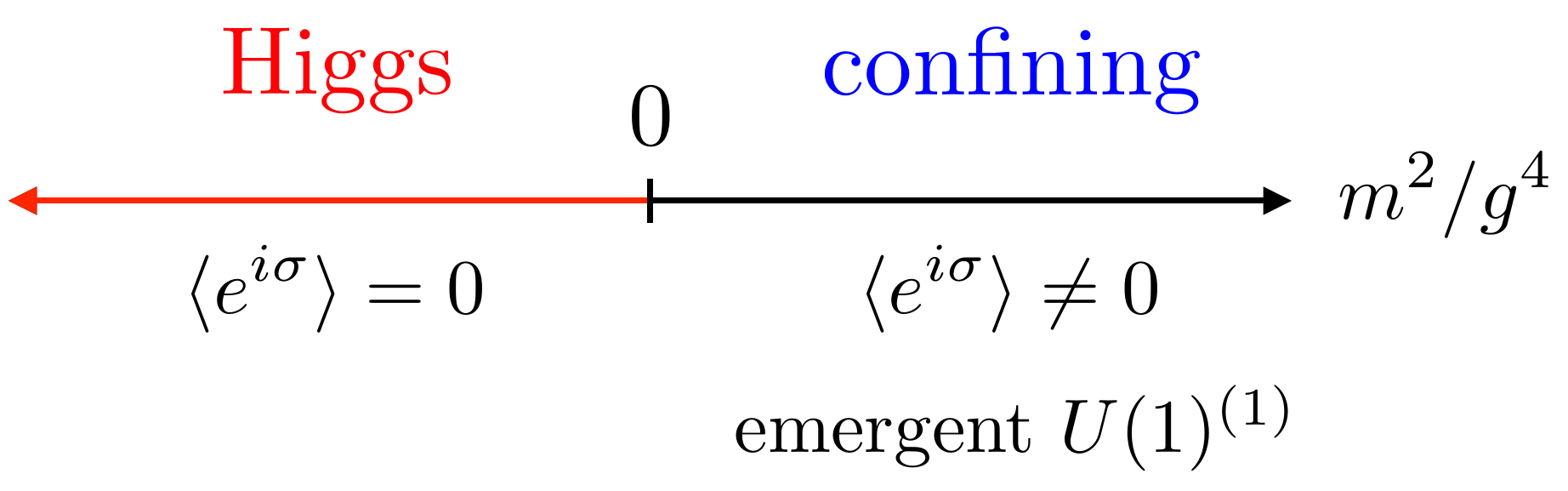}
  \caption{A sketch of the minimal phase diagram of 3d scalar QED, which has Higgs and confining phases which are separated by at least one phase boundary where the realization of the $U(1)_m^{(0)}$ symmetry changes and  the expectation values of monopole operators $e^{i\sigma}$ are non-analytic.  The confining phase is characterized by an emergent $U(1)^{(1)}_e$ symmetry.}
  \label{fig:higgs_confinement_phase_diagram}
\end{figure}

One interesting consequence of the emergent $U(1)^{(1)}_e$
symmetry of 3d massive scalar QED with a $U(1)^{(0)}_m$ magnetic symmetry
involves its ground state structure.  While naively the explicit breaking of the
$U(1)^{(1)}_e$ kills the mixed 't Hooft anomaly, the anomaly manages to maintain
its grip on the ground state structure thanks to  Eq.~\eqref{eq:anomaly_correlators}
and the discussion in the preceding section.  The non-triviality of
Eq.~\eqref{eq:anomaly_correlators}  at long distances means that when $m$ is finite
(but sufficiently positive) the ground state must remain non-trivial. This is in accordance with Ref.~\cite{Delacretaz:2019brr} which argues that the gapless mode in this regime is a direct consequence of an emergent 't Hooft anomaly. On the
other hand if $m^2$ is sufficiently negative, Eq.~\eqref{eq:anomaly_correlators}
trivializes.  As a result, the ground state can be (and is) trivially gapped, as can be verified by
a standard semiclassical calculation, see Fig.~\ref{fig:higgs_confinement_phase_diagram} for an illustration.

If we had preserved the $U(1)^{(1)}_e$ symmetry but explicitly broken
$U(1)^{(0)}_m$ by adding unit-charge monopole operators to the action, we would
have landed on Eq.~\eqref{eq:linear_conf_action}, a QFT with area-law-like
`confinement' (in the same loose sense that QCD with finite mass quarks is
confining)  and a trivially gapped ground state.   This is another illustration
of the
robustness~\cite{Wen:1991rp,Vestergren_2005,Hastings_2005,
Iqbal:2021rkn,Cordova:2022rer,Pace:2022cnh,Pace:2023gye,Armas:2023tyx,Verresen:2022mcr,Thorngren:2023ple}
of $1$-form symmetries compared to $0$-form symmetries. Even when $1$-form
symmetries are explicitly broken, they can enforce interesting constraints on
the low-energy physics, in contrast to $0$-form symmetries.  

Another interesting byproduct of the discussion above has to do with
Higgs-confinement continuity, namely the common lore that the Higgs and
confining regimes of gauge theories with fundamental-representation Higgs fields
should be smoothly connected. Higgs-confinement continuity was proven in some
lattice models in Refs.~\cite{Fradkin:1978dv,Banks:1979fi}, where a key assumption was
the absence of any global symmetries under which the Higgs field is charged. The
heuristic argument is that in such a situation the Higgs and confining regimes
are smoothly connected because there are no order parameters that could
distinguish them. The model we have discussed above in
Eq.~\eqref{eq:charge1higgs} provides an immediate and non-trivial
counter-example to the popular interpretation of the results of
~\cite{Fradkin:1978dv,Banks:1979fi} as implying Higgs-confinement continuity
even outside the context of the specific models analyzed in those references. 

The model in Eq.~\eqref{eq:charge1higgs} has a unit-charge Higgs field $\phi$
and a $U(1)^{(0)}_m$ global symmetry under which the Higgs field is
\emph{neutral}. The Higgs and confining phases are separated by at least one
phase transition where the realization of $U(1)^{(0)}_m$ changes, and our
results imply that the `confining' phase can be \emph{defined} as the phase with
an emergent $1$-form symmetry, which is separated by at least one phase boundary
from the `deconfined' Higgs phase. What makes this counter-example to the standard lore non-trivial
is that the Higgs field $\phi$ is not charged under any global symmetry, yet its
`condensation' controls the emergent mixed anomaly and subsequent breaking of
$U(1)^{(0)}_m$.  Other possible non-trivial counter-examples to
Higgs-confinement continuity in some systems with global symmetries were
recently discussed in e.g. Refs.~\cite{Cherman:2018jir,Cherman:2020hbe}.

\section{Conclusions} 

We have put forth necessary conditions for the emergence of
1-form symmetries at long distances. The key requirement is that the
long-distance correlation functions of operators be topological and non-trivial.
By this criterion, a 1-form symmetry which is not spontaneously broken nor
involved in 't Hooft anomalies cannot emerge at long distances when broken
explicitly by finite-mass charged matter. Our definition agrees with the usual
lore that the implications of spontaneously-broken higher-form symmetries are
robust against explicit breaking.  Moreover, the implications of a higher-form
symmetry can remain robust against explicit breaking even when the symmetry is
\emph{not} spontaneously broken when the symmetry is involved in an 't Hooft
anomaly.

While for simplicity we focused on 1-form symmetries in 3d, our definition can
 in principle be applied to higher-form symmetries in any dimension. It would be
particularly interesting to analyze a wider class of 't Hooft anomalies that
arise in situations where the higher-form symmetries are not spontaneously
broken. For instance, 4d $SU(N)$ $\mathcal{N}=1$ super-YM theory has a
$\mathbb{Z}_N$ $1$-form symmetry and a $\mathbb{Z}_{N}$ $0$-form chiral symmetry with a
mixed 't Hooft anomaly~\cite{Gaiotto:2014kfa,Komargodski:2017smk}. If we add a
massive fundamental-representation fermion, do we expect an emergent 1-form
symmetry at long distances? The arguments in Sec.~\ref{sec:anomaly} and Sec.~\ref{sec:consequences} can be applied to the line-like intersections of Gukov-Witten surface operators with codimension-1 operators generating the chiral symmetry --- while ordinary Wilson lines are confined in $\mathcal{N}=1$ SYM, these non-genuine lines will be approximately topological at long distances and contribute to non-trivial correlation functions. 

It may also be worthwhile to explore the connection of our emergence criterion
with the phenomenon of symmetry
fractionalization, see e.g. Refs.~\cite{Jackiw:1975fn,PhysRevB.62.7850,PhysRevB.100.115147,
PhysRevX.5.041013,Tarantino2016,CHEN20173,Brennan:2022tyl,Delmastro:2022pfo},
where the charged particles breaking the 1-form symmetry transform in projective
representations of a 0-form global symmetry.

Finally, throughout this paper, we focused on the interplay of the large-mass
limit, in which the symmetry-breaking matter field contributions are clearly
suppressed, and the large-distance limit, where it is less obvious that symmetry breaking effects are suppressed.  We should mention that $SU(N)$ QCD has another limit where matter field contributions are
clearly suppressed, namely the large $N$ limit where the  't Hooft coupling,
number of flavors $N_f$ and mass parameters are fixed as $N \to \infty$.  Does a $1$-form symmetry
emerge in the large $N$ limit? We recently studied this question together with
M. Neuzil in Ref.~\cite{Cherman:2022eml}. It is well-known that screening of
large Wilson loops is $1/N$ suppressed so long as the large $N$ limit is taken
before the large-loop limit, so the specific failure mechanism for the emergence
of a $1$-form symmetry discussed in the present paper does not apply. However, a
$1$-form symmetry fails to emerge anyway, because the $U_{\alpha}(\Gamma)$
operators do not become topological when $N\to \infty$.  Their expectation
values take the form of Eq.~\eqref{eq:U_perimeter} with $\mu_{\alpha} \sim N$, so
that $\langle U_{\alpha}(\Gamma) \rangle = 0$ at large $N$. It remains an
interesting open question  whether there is some notion of an `emergent
symmetry' which would explain the Wilson loop selection rules of large $N$ QCD.

\section*{Acknowledgements}

We are grateful to Fiona Burnell, Clay Cordova, Jacques Distler, Thomas Dumitrescu,
Po-Shen Hsin, Nabil Iqbal, Jake McNamara, Pierluigi Niro, Srimoyee Sen, Shu-Heng Shao and Laurence Yaffe for
discussions. A.~C. and T. J. are supported in part by the Simons Foundation Grant No.
994302. T.J. also acknowledges support from a Doctoral Dissertation Fellowship at the
University of Minnesota. We thank the Institute for Nuclear Theory at the
University of Washington for its kind hospitality and stimulating research
environment during our participation in the INT Workshop  INT-21R-1A, so that we were also supported in part by the INT's U.S. Department of Energy grant No.
DE-FG02- 00ER41132.  We are also grateful to the participants of the workshop ``Traversing the Particle Physics Peaks - Phenomenology to Formal" at the Aspen Center for Physics for valuable discussions,  which were supported by National Science Foundation grant PHY-2210452.


\appendix

\section{$U(1)^{(0)} \times U(1)^{(1)}$ 't Hooft anomaly}
\label{sec:anomaly_done_right}

We give a more precise characterization of the anomaly described in the main
text. This allows us to identify operators with non-trivial correlation
functions which would not be obvious in a naive continuum approach. 

We follow the formalism of Ref.~\cite{Alvarez:1984es}, see
also~\cite{Cordova:2019jnf}, where we describe gauge fields in terms of
patches, transition functions, and cocycle conditions. We choose an open cover
$\{ U_I \}$ of the spacetime manifold.  Then we pick an associated partition of
the manifold into three-dimensional closed regions $\{\sigma_I\}$ with $\sigma_I
\subset U_I$, such that $\sigma_{IJ} = \sigma_I \cap \sigma_J \subset U_I \cap
U_J$ are 2-dimensional and contained in double overlaps, $\sigma_{IJK} =
\sigma_I \cap \sigma_J \cap \sigma_K \subset U_I \cap U_J \cap U_K$ are
one-dimensional and contained in triple overlaps, and $\sigma_{IJKL} = \sigma_I
\cap \sigma_J \cap \sigma_K \cap \sigma_L \subset U_I \cap U_J \cap U_K \cap
U_L$ are points contained in quadruple overlaps. 

The dynamical $U(1)$ gauge field is described by a collection of fields
$(a^\1_I, a^\0_{IJ}, a^\m1_{IJK})$. The starting point is an $\RR$-valued 1-form
gauge field on each patch $a_I^{(1)}$. In the language of
Ref.~\cite{Alvarez:1984es} this is a 1-form-valued 0-cochain. We use a shorthand
notation $\delta$ to take differentials of fields on various overlaps: for
instance $(\delta a^\1)_{IJ} = a^\1_J - a^\1_I$. On double overlaps have
$(\delta a^\1)_{IJ} = da^\0_{IJ}$ where $a^\0_{IJ} \in \RR$ is a real-valued,
0-form transition function. Since $a^\0$ has two `patch' indices it is a
1-cochain (with values in 0-forms). It satisfies $(\delta a^\0)_{IJK} =
a^\0_{JK} - a^\0_{IK} + a^\0_{IJ}= 2\pi a^\m1_{IJK}$ with $a^\m1_{IJK} \in \ZZ$
is a constant integer defined on triple-overlaps (an integer 2-cochain). The
$\delta$ operation (which is the coboundary operator) squares to 0 ($\delta^2 =
0$ and commutes with $d$, so that e.g. under an ordinary gauge transformation
$(\delta a^\1)_{IJ} \to (\delta a^\1)_{IJ} + (\delta d\lambda^\0)_{IJ} = (\delta
a^\1)_{IJ} + d (\delta\lambda^\0)_{IJ}$. To summarize, the set of relations is
\begin{subequations}
  \begin{align}
  (\delta a^\1)_{IJ} &= da^\0_{IJ}, \\
  (\delta a^\0)_{IJK} &= 2\pi\, a^\m1_{IJK}, \\
 (\delta a^\m1)_{IJKL} &= 0 \,,
\end{align}
\label{eq:a_relations} 
\end{subequations}
\!\!\!where in the last equation we are looking at quadruple overlaps, and the `cocycle condition' $\delta a^\m1 = 0$ means there are no magnetic monopoles. The full gauge
redundancy is 
\begin{subequations}
  \begin{align}
  a^\1_I &\to a^\1_I + d\lambda^\0_I, \\
  a^\0_{IJ} &\to a^\0_{IJ} + (\delta\lambda^\0)_{IJ} + 2\pi\, m^\m1_{IJ}, \\
  a^\m1_{IJK} &\to a^\m1_{IJK} + (\delta m^\m1)_{IJK}\,,
  \end{align}
\end{subequations}
where the function $\lambda_I^\0$ and constants $m_{IJ}^\m1 \in \ZZ$ represent
small and large gauge transformations, respectively. 

We take the same set of data for the background 1-form $U(1)$ gauge field,
$(A_I^\1, A_{IJ}^\0, A^\m1_{IJK})$, where $A^\1$ is a 1-form, $A^\0$ is a
0-form, $A^\m1$ is a constant integer, and $\Lambda^\0_I$ and
$M^\m1_{IJ}$ will denote the small and large background gauge transformations.
In this formalism, a symmetry operator supported on a surface is described by
$A_I^\1 = 0$ everywhere and $A^\0_{IJ}$ constant on a set of double-overlaps, so that the corresponding set of $\sigma_{IJ}$'s constitute the surfaces on which the
defect is supported. The surface may have a boundary as long as $\delta A^\0$ a
multiple of $2\pi$, with $A^\m1_{IJK}$ activated on the boundary.   The
well-known statement that a \emph{generic} $U(1)$ background gauge field cannot be
expressed in terms of a network of symmetry defects is reflected in the fact
that it is possible to activate a non-zero 1-form $A_I^\1$ on each patch.

Up to signs, the precise version of $\frac{i}{2\pi} \int A \wedge da$ is
\begin{multline} \label{eq:Ada} 
`` \frac{i}{2\pi} \int A \wedge da " = \frac{i}{2\pi}\sum_I \int_{\sigma_I} A_I^\1 \wedge da_I^\1 \\
+ \frac{i}{2\pi} \sum_{I<J}\int_{\sigma_{IJ}} A^\0_{IJ} \wedge da^\1_J - \frac{i}{2\pi} \sum_{I<J<K} A^\m1_{IJK} \int_{\sigma_{IJK}} a_K^\1 \\
- i \sum_{I<J<K<L} \left. A^\m1_{IJK}\,a^\0_{KL} \right|_{\sigma_{IJKL}}\,.
\end{multline}
It is invariant under gauge transformations of all fields and is well-defined.
What happens if we turn on  $A^\m1_{IJK} = q$ on some triple overlap?  First, we see
that this inserts a Wilson line, and the cocycle condition on $A^\m1$ implies
that this line must be closed. Second, the relations between $A^{\m1}_{IJK}$ and
neighboring $A^{\0}_{IJ}$ implies that we must also turn on $A^\0_{IJ} =
2\pi q$ on some neighboring double overlap, which inserts a field-strength surface
operator involving $da^\1$ on that overlap.   This gives a rigorous definition of 
the operator $V_{2\pi q}(D)$ in Eq.~\eqref{eq:anomaly_correlators} the main text.

The 2-form gauge field is
described by $(B_I^\2, B^\1_{IJ}, B^\0_{IJK}, B^\m1_{IJKL})$. Here
$B^\2$ is a real 2-form on each patch, $B^\1$ is a real 1-form on double
overlaps, $B^\0$ is a real 0-form on triple overlaps, and $B^\m1$ is a constant
integer on quadruple overlaps. They are related via 
\begin{subequations}
\begin{align}
(\delta B^\2)_{IJ} &= dB^\1_{IJ}, \\
 (\delta B^\1)_{IJK} &= dB^\0_{IJK},\\
(\delta B^\0)_{IJKL} &= 2\pi\, B^\m1_{IJKL}\,. 
\end{align}
\end{subequations}
Under gauge transformations we have 
\begin{subequations}
\begin{align}
B^\2_{I} &\to B^\2_{I} + d\Pi^\1_{I}, \\
 B^\1_{IJ} &\to B^\1_{IJ} + (\delta\Pi^\1)_{IJ} + d\Pi^\0_{IJ}, \\
 B^\0_{IJK} &\to B^\0_{IJK} + (\delta\Pi^0)_{IJK} + 2\pi L^\m1_{IJK}, \\
  B^\m1_{IJKL} &\to B^\m1_{IJKL} + (\delta L^\m1)_{IJKL}\,,
\end{align} 
\end{subequations}
where $\Pi^\1_I$ is a real 1-form, $\Pi^0_{IJ}$ is a real 0-form, $L^\m1_{IJK}$
is a constant integer. Again, a generic 2-form $U(1)$ gauge field cannot be
associated with a network of symmetry defects, but if $B^\2 =0, B^\1 = 0$ we can
think of $B^\0$ activated on a set of triple-overlaps as inserting a
codimension-2 symmetry defect on lines (corresponding to a set of
$\sigma_{IJK}$'s). 

The dynamical gauge field $a^\1$ shifts under background 1-form gauge
transformations. Accordingly, in the presence of the background field for the
1-form symmetry the relations Eq.~\eqref{eq:a_relations} are modified to 
\begin{subequations}
\begin{align}
(\delta a^\1)_{IJ} &= da^\0_{IJ} + B^\1_{IJ}, \\
  (\delta a^\0)_{IJK} &= 2\pi a^\m1_{IJK} -B^\0_{IJK}, \\
   (\delta a^\m1)_{IJKL} &= B^\m1_{IJKL}\,.
\end{align}
\end{subequations}
These relations are invariant under the background gauge transformations
\begin{subequations}
\begin{align}
a^\1_I &\to a^\1_I + \Pi^\1_I ,\\
 a^\0_{IJ} &\to a^\0_{IJ} - \Pi^\0_{IJ}, \\
a^\m1_{IJK} &\to a^\m1_{IJK} + L^\m1_{IJK}\,.
\label{eq:modified_cocycle}
\end{align}
\end{subequations}
The anomalous shift of Eq.~\eqref{eq:Ada} under 1-form background gauge
transformations is then 
\begin{multline} \label{eq:shift_1} 
\frac{i}{2\pi}\sum_I \int_{\sigma_I} A^\1_I \wedge d\Pi^\1_I + \frac{i}{2\pi} \sum_{I<J}\int_{\sigma_{IJ}} A^\0_{IJ} \wedge d\Pi^\1_J \\
- \frac{i}{2\pi} \sum_{I<J<K} A^\m1_{IJK} \int_{\sigma_{IJK}} \Pi^\1_K \\
+ i \sum_{I<J<K<L} \left. A^\m1_{IJK}\,\Pi^\0_{KL} \right|_{\sigma_{IJKL}}\,,
\end{multline} 
As we mentioned above,  if we insert $V_{2\pi q}(D)$ by taking $A^\0_{IJ} = 2\pi q$
on a set of double-overlaps tiling $D$, we must also have $A^\m1_{IJK} =\pm q$
on the set of triple overlaps containing the boundary of $D$.  We now insert $U_{\beta}(\Gamma)$ by
setting  $B^\0_{IJK}$ equal to a constant $\beta$ on the triple overlaps tiling
the closed loop $\Gamma$, which we assume pierces $D$. Removing this curve with
an appropriate 1-form gauge transformation with $\Pi^\0 = \beta$, we pick up a
phase $e^{i\beta q}$ from the last term above.   This reproduces
Eq.~\eqref{eq:surface_anomaly} in the main text, except that we have not yet
discussed the counter-term dependence.

We also have anomalous shifts of Eq.~\eqref{eq:Ada} involving 0-form gauge
transformations of $A$ coming from the modified cocycle condition
Eq.~\eqref{eq:modified_cocycle}, 
\begin{multline} \label{eq:shift_2}
-\frac{i}{2\pi} \sum_{I<J} \int_{\sigma_{IJ}} \Lambda^\0_I \wedge dB^\1_{IJ} -i \sum_{I<J<K} M^\m1_{IJ}\int_{\sigma_{IJK}} B^\1_{JK} \\
+ i \sum_{I<J<K<L} \left. M^\m1_{IJ} \, B^\0_{JKL} \right|_{\sigma_{IJKL}}\,.
\end{multline}

We can now contemplate adding various counter-terms involving $B$. In particular
if we add
\begin{align}
&`` \frac{-i}{2\pi} \int A \wedge B "  = -\frac{i}{2\pi}\sum_I \int_{\sigma_I} A_I^\1 \wedge B_I^\2  \\
&- \frac{i}{2\pi}\sum_{I<J}\int_{\sigma_{IJ}} A^\0_{IJ} \wedge B^\2_J 
-\frac{i}{2\pi}\sum_{I<J<K<L} \left. A^\0_{IJ} \, B^\0_{JKL}\right|_{\sigma_{IJKL}}\,, \nonumber
\end{align}
then we can cancel all the terms in Eq.~\eqref{eq:shift_2} and the first two
terms in Eq.~\eqref{eq:shift_1}. This is equivalent to setting $b_{\text{CT}} = 1$ in Eq.~Eq.~\eqref{eq:naive_anomaly}. We then have
\begin{align}
&`` \frac{i}{2\pi} \int A \wedge (da-B) " = \frac{i}{2\pi}\sum_I \int_{\sigma_I} A^\1_I \wedge (da^\1 -B^\2)_I\nonumber \\
&+ \frac{i}{2\pi} \sum_{I<J}\int_{\sigma_{IJ}} A^\0_{IJ} \wedge (da^\1-B^\2)_J\nonumber \\
&- \frac{i}{2\pi} \sum_{I<J<K} A^\m1_{IJK} \int_{\sigma_{IJK}} a^\1_K\nonumber \\
&- i \sum_{I<J<K<L} \left. A^\m1_{IJK}\, a^\0_{KL} \right|_{\sigma_{IJKL}}\nonumber \\
&-\frac{i}{2\pi}\sum_{I<J<K<L} \left. A^\0_{IJ} \, B^\0_{JKL}\right|_{\sigma_{IJKL}} \,,
\label{eq:A_daB}
\end{align}
The anomalous shift of Eq.~\eqref{eq:A_daB} is 
\begin{align} \label{eq:shift_3} 
&S_{\text{anomaly}, b_{\text{CT}}=1} \\
&= -i \sum_{I<J} M^\m1_{IJ} \int_{\sigma_{IJ}} B^\2_J -\frac{i}{2\pi}\sum_{I<J<K} A^\m1_{IJK} \int_{\sigma_{IJK}} \Pi^\1_K \nonumber  \\
&+ i \sum_{I<J<K<L}\left. A^\m1_{IJK}\,\Pi^\0_{KL} \right|_{\sigma_{IJKL}}\nonumber \\
&- \frac{i}{2\pi} \sum_{I<J<K<L}\left. A^\0_{IJ} (\delta\Pi^\0 + 2\pi L^\m1)_{JKL}\right|_{\sigma_{IJKL}}\nonumber\\
 &-i \sum_{I<J<K<L}\left. M^\m1_{IJ}\, (\delta\Pi^\0)_{JKL} \right|_{\sigma_{IJKL}}\nonumber \\
 &-\frac{i}{2\pi}\sum_{I<J<K<L}\left. (\delta\Lambda^\0)_{IJ}\,B^\0_{JKL}\right|_{\sigma_{IJKL}}\nonumber
\end{align}
Now if we consider the same configuration from before with $A^\0_{IJ} = 2\pi q,
A^\m1_{IJK} = q$, the two terms above in the second and third line cancel and
$\langle V_{2\pi q }(D) U_{\beta}(\Gamma) \rangle$ has no non-trivial phases.

However, we can consider a configuration where $B^\0_{IJK} = 2\pi k $ on a set of
triple-overlaps containing an open line $L$, with $B^\m1_{IJKL} = \pm k$ at the
quadruple overlaps containing the endpoints of $L$. This represents the
insertion of an open generator of the $2\pi$ 1-form transformation $U_{2\pi k}(L)$
(which would be trivial absent an anomaly). If we also insert a 0-form generator
$V_{\alpha}(\Sigma)$ on a closed surface $\Sigma$ enclosing the endpoint, with
$A^\0_{IJ} = \alpha$ for some constant $\alpha$ on an appropriate set of
double-overlaps containing $\Sigma$, then the gauge transformation $A^\0_{IJ}
\to A^\0_{IJ} + (\delta\Lambda^\0)_{IJ}$ used to unlink the surface $\Sigma$
from the the endpoint of the 1-form generator $U_{2\pi k}(L)$ will yield the phase
$e^{i\alpha k}$ from the last term in Eq.~\eqref{eq:shift_3}.  This reproduces
Eq.~\eqref{eq:endpoint_anomaly} for the behavior of $\langle V_{\alpha}(\Sigma)
U_{2\pi k}(L) \rangle$.

\bibliography{non_inv}

\end{document}